\documentclass[12pt]{article}
\usepackage{multicol}
\usepackage[round]{natbib}
\usepackage{amssymb}
\usepackage{geometry} 
\usepackage{indentfirst}
\usepackage[fleqn]{amsmath}
\usepackage{amsmath}
\usepackage{bm}
\usepackage{caption}
\usepackage{listings, xcolor}
\usepackage{verbatim}
\usepackage{graphicx} 
\usepackage{float} 
\usepackage{subfigure} 
\usepackage[justification=centering]{caption}
\usepackage{setspace} 
\usepackage[
bookmarks=true,
colorlinks=true,
linkcolor=blue,
urlcolor=blue,
citecolor=blue,
pdftex,
bookmarks=true,
linktocpage=true,   
hyperindex=true
]{hyperref}
{ \title{Bayesian Poisson Log-normal Model with Regularized
Time Structure for Mortality Projection of Multi-population}   
 \date{}                  
\author{Zhen Liu$^1$, Xiaoqian Sun$^1$, Leping Liu$^2$\thanks{Part of his research has been supported by National Science Foundation of China (NSFC) under Grant \#71771163.}, Yu-Bo Wang$^1$\thanks{Corresponding author: yubow@clemson.edu}\\
$^1$School of Mathematical and Statistical Sciences, \\Clemson University, SC, USA\\
$^2$Department of Statistics, Tianjin University of \\Finance and Economics, Tianjin, China}

\begin{document}
\maketitle                                	
\newpage
\renewcommand{\baselinestretch}{1.8} \normalsize
\newcommand{\age}{{\mbox{\tiny age}}}
\newcommand{\tim}{{\mbox{\tiny time}}}
\newcommand{\var}{{\mbox{Var}}}
\newcommand{\E}{\mbox{E}}
\newcommand{\nn}{\nonumber}

\newcommand{\bw}{{\mbox{\boldmath $w$}}}
\newcommand{\bW}{{\mbox{\boldmath $W$}}}
\newcommand{\bQ}{{\mbox{\boldmath $Q$}}}

\newcommand{\balpha}{{\mbox{\boldmath $\alpha$}}}
\newcommand{\bbeta}{{\mbox{\boldmath $\beta$}}}
\newcommand{\bkappa}{{\mbox{\boldmath $\kappa$}}}
\newcommand{\btheta}{{\mbox{\boldmath $\theta$}}}
\newcommand{\bvarphi}{{\mbox{\boldmath $\varphi$}}}
\newcommand{\bnu}{{\mbox{\boldmath $\nu$}}}
\newcommand{\bSigma}{{\mbox{\boldmath $\Sigma$}}}

\centerline{Abstract}	
The improvement of mortality projection is a pivotal topic in the diverse branches related to insurance, demography, and public policy. Motivated by the thread of Lee-Carter related models,
we propose a Bayesian model to estimate and predict mortality rates for multi-population. This new model features in information borrowing among populations and properly reflecting variations of data. It also provides a solution to  a long-time overlooked problem: model selection for dependence structures of population-specific time parameters. By introducing a dirac spike function,  simultaneous model selection and estimation for population-specific time effects can be achieved without much extra computation cost. We use the Japanese mortality data from Human Mortality Database to illustrate the desirable properties of our model.
 

\noindent {Keywords:} Bayesian approach; Poisson LC model; Dirac spike; Mortality projection.
	
\section{Introduction}

Mortality projection has become an important topic in demographics since it is greatly involved in many policy makings including but not limited to public health, pension, retirement system and labor resources. Especially for those developed and developing countries that experience population aging due to rapid growth of life expectancy and decline of mortality after 1950s (\citealp{tuljapurkar2000universal}), a thorough and well established policy relies on an accurate prediction of mortality trajectory. 


Over last decades, stochastic models have been widely applied to mortality projection because the produced forecasts along with intervals can properly capture uncertainties over time and inform decision makings. The Lee-Carter (LC) model, a leading model proposed by \cite{lee1992modeling}, decomposes the centered mortality force in log scale as the product of age and time effects, and considers a random walk with drift (RWD) model on time effect profile for the prediction purpose. This log-bilinear model was first developed for the U.S. mortality data from 1933 to 1987, and now becomes a benchmark widely implemented in all-cause or cause-specific mortality data.     
Following this structure, \cite{brouhns2002poisson} proposed the Poisson model for the number of deaths instead of directly modeling the observed mortality rate. Although it may encounter overdispersion due to the limitation of a Possion distribution, the Poisson LC model distinguishes the cases with the same observed rate but different exposures at risk, and hence, takes advantage of more information from the data. 
On this basis, \cite{czado2005bayesian} extended to a Bayesian framework to bypass the two-stage estimation procedure while preserving uncertainty from the model in the posterior predictive distributions of mortality rates. \cite{wong2018bayesian} further considered a random effect to accommodate overdispersion. Other related works can be referred to \cite{girosi2003demographic}, \cite{renshaw2003lee}, \cite{cairns2006two} and \cite{plat2009stochastic}.

Motivated by the benefits of borrowing information among populations, many works, such as \cite{li2005coherent}, \cite{cairns2011bayesian}, \cite{li2011measuring}, and \cite{antonio2015bayesian}, have focused on simultaneously projecting the mortality rates of multiple groups by encapsulating the common and population-specific age and time components in the models. In this paper, we revisit the works of \cite{wong2018bayesian} and \cite{antonio2015bayesian}, and develop a new multi-population model that can entertain the potential overdispersion in data. We also relax the model assumption on each population-specific time component by considering an autoregressive model of order one (AR(1)) with a drift, where the drift parameter and the slope associated with time are hierachically regulated by a dirac spike and slab prior \citep{george1993variable,ishwaran2005spike, malsiner2018comparing}. As a result, model selection between AR(1) with and without a drift on population-specific time components is conducted simultaneously with estimation and prediction.




The remainder of the paper is organized as follows. In Section 2, we review the Lee-Carter model and its recent developments. Section 3 introduces the proposed model along with 
 the prior settings and detailed steps of  an Markov chain Monte Carlo (MCMC) sampling. In Section 4, the proposed method is applied to Japan gender-specific mortality data 
 between 1951 and 2016 from the Human Mortality Database (HMD). 
 To evaluate its performance, the results based on the model by \cite{antonio2015bayesian} are also included as a comparison. 
 Finally, we conclude with a discussion in Section 5.

\section{Lee-Carter Model and its Extensions}

 \cite{lee1992modeling} introduced a stochastic model for modeling the US mortality data from 1933 to 1987 in an attempt to forecast the future mortality rate during 1988-2065. Suppose $\Theta_{\age}=\{x_1,x_1+1,\dots,x_1+M-1\}\equiv\{x_1,x_2,\dots,x_M\}$ and $\Theta_{\tim}=\{t_1,t_1+1,\dots, t_1+N-1\}\equiv\{t_1,t_2,\dots,t_N\}$ 
 denote the sets of age and time considered in the training dataset
, respectively, the Lee-Carter model is then given by
\begin{equation}
\log m_{x,t}=\alpha_x+\beta_x\kappa_t+\epsilon_{x,t}, \label{eq1}
\end{equation}
where $m_{x,t}$ is the observed mortality rate for the group aged $x$ at time $t$, 
$\epsilon_{x,t}$ is the error term, and $x\in \Theta_{\age}$ and $t\in \Theta_{\tim}$. Essentially, this model is a special case of log-linear model in a cross table because 
$\log m_{x,t}$ is decomposed as the product of age ($\beta_x$) and time ($\kappa_t$) effects plus an age-specific intercept ($\alpha_x$), where  $\beta_x$ is a constant over time while an additional time series model is placed on $\kappa_t$ for the prediction purpose.  To make $\alpha_x$, $\beta_x$, and $\kappa_t$ in (\ref{eq1}) estimable, two constraints are imposed in \cite{lee1992modeling}: $\sum_{x\in \Theta_{\age}}\beta_x=1$ and $\sum_{t\in \Theta_{\tim}}\kappa_t=0$. With such constraints, the age-specific intercept $\alpha_x$ is first estimated as the mean of log rates at age $x$ observed across time, 
and then the singular value decomposition (SVD) is applied to the matrix of centered log rates, $\log m_{x,t}-\hat{\alpha}_x$, to estimate  ${\beta}_x$ and ${\kappa}_t$.
Based on $\{\hat{\kappa}_t, \mbox{for } t\in\Theta_{\tim}\}$, the autoregressive integrated moving average (ARIMA) model is separately fitted to forecast the future time components $\kappa_t$ and thus the mortality projection for any future year can be obtained.  




Considering additional information contained in the exposure at risk ($E_{x,t}$), \citet{brouhns2002poisson} modified the LC model into the following Poisson framework 
\begin{align}
D_{x,t}|\mu_{x,t} \sim \text{Poisson}(E_{x,t}\mu_{x,t}) \quad \text{with} \quad \log\mu_{x,t}=\alpha_x+\beta_x\kappa_t,  \label{eq2}
\end{align}
where $D_{x,t}$ is the death toll for the group aged $x$ at time $t$, and $\mu_{x,t}$ is the corresponding theoretic mortality rate. Note that $\mu_{x,t}$ differs from $m_{x,t}=D_{x,t}/E_{x,t}$ in (\ref{eq1}), and that the cases with the same observed rate will have different likelihood values if their $E_{x,t}$s' differ. With the same constraints on $\beta_x$ and $\kappa_t$, \cite{brouhns2002poisson} adopted the maximum likelihood estimation for  $\alpha_x$, $\beta_x$ and $\kappa_t$ in (\ref{eq2}), and similarly, fitted $\{\hat{\kappa}_t, \mbox{for } t\in\Theta_{\tim}\}$ with the ARIMA model afterwards. 

It is clear that both the LC and Poisson LC models are two-stage analyses, where the main model (that is, (\ref{eq1}) or (\ref{eq2})) and the ARIMA model are fitted for estimation and prediction, respectively. Consequently, it may underestimate the uncertainty of the mortality projection. To properly reflect the uncertainty from the estimation process in the main model into forecasting, \cite{czado2005bayesian} considered the Poisson LC model in Bayesian framework, where an MCMC sample is drawn from the posterior distribution of the joint model and used to construct the posterior predictive distribution of mortality rates in the future. 
Another efforts on improving the Poisson LC model can be found in \cite{wong2018bayesian}, where the proposed method tackles with overdispersion potentially encountered in the Poisson model. Letting $\nu_{x,t}$ denote a random effect following $N(0,\sigma^2)$, the normal distribution with mean 0 and variance $\sigma^2$, they proposed the Poisson log-normal Lee-Carter (PLNLC) model as 
\begin{align}
D_{x,t}|\mu_{x,t} \sim \text{Poisson}(E_{x,t}\mu_{x,t}) \quad \text{with} \quad \log\mu_{x,t}=\alpha_x+\beta_x\kappa_t+\nu_{x,t}.  \label{eq3}
\end{align}
With this additional diffusion $\nu_{x,t}$, the PLNLC model relaxes the equality constraint on mean and variance as follows
\begin{align*}
&\E[D_{x,t}]=\E[\E(D_{x,t}|\nu_{x,t})]=E_{x,t}\exp(\alpha_x+\beta_x\kappa_t+\frac{1}{2}\sigma^{2}),\\
&\var[D_{x,t}]=\E[\var(D_{x,t}|\nu_{x,t})]+\var[\E(D_{x,t}|\nu_{x,t})]\\
&\hspace{1.6cm}=\E[D_{x,t}]\times\{1+\E[D_{x,t}]\times[\exp(\sigma^2)-1]\}\geq \E[D_{x,t}],
\end{align*}
and hence, has a wider application in mortality data.

Besides, inspired from \cite{li2005coherent} and \cite{renshaw2003lee}, the works considering two bilinear terms, \cite{antonio2015bayesian}  extended (\ref{eq2}) to the following Poisson log-bilinear model for a $n$-population data set
\begin{align}
D_{x,t}^{(i)}|\mu_{x,t}^{(i)} \sim \text{Poisson}(E_{x,t}^{(i)}\mu_{x,t}^{(i)}) \quad \text{with} \quad \log\mu_{x,t}^{(i)}=\alpha_x^{(i)}+\beta_x\kappa_t+\beta_x^{(i)}\kappa_t^{(i)},  \label{eq4}
\end{align}   
where the first bilinear term $\beta_x\kappa_t$ now denotes the overall effect shared by all populations aged $x$ at time $t$, and the superscript $(i)$ marks $i^{th}$ population-specific term so  $\alpha_x^{(i)}$ and $\beta_x^{(i)}\kappa_t^{(i)}$, for $i=1,2,,\dots,n$, are a population-specific intercept and effect, respectively.                                                                                                                                                                                                                       To identify (\ref{eq4}), additional constraints on population-specific age and time effects are required: $\left\|\beta_{x}^{(i)}\right\|_2 =1$ and $\sum_{t\in\Theta_{\tim}}\kappa_{t}^{(i)}=0$, where $\left\| . \right\|_2$ represents the $L_2$ norm of a vector.  Through jointly investigating related populations, (\ref{eq4}) tends to be more efficient than the separate modeling using PLC on each population.

In this paper, we consider pros and cons of the works mentioned above, and propose the Poisson Log-normal model for mortality projection of multi-population in the Bayesian framework. This new model not merely combines the PLNLC model with (\ref{eq4}), but also simultaneously conducts model selection of time structures of $\kappa_t^{(i)}$ via a dirac spike and slab prior.  As a result, it can serve for more varieties of mortality data.  We introduce our model formally in Section 3.

\section{The Proposed Model}

\subsection{Bayesian Poisson Log-normal Lee-Carter Model with Regularized Time Structure for Multi-population}

Let $\nu_{x,t}^{(i)}$ denote the $i^{th}$ population-specific random effect following $N(0, \sigma_{i}^{2})$ for $i=1,2,\dots,n$. We propose the Bayesian Poisson log-normal  Lee-Carter model 
for n-population (BPLNLCrm) as follows
\begin{align}
&D_{x,t}^{(i)}|\mu_{x,t}^{(i)} \sim \text{Poisson}(E_{x,t}^{(i)} \mu_{x,t}^{(i)})  \quad \text{with} \quad \log \mu_{x,t}^{(i)}= \alpha_x^{(i)}+\beta_x\kappa_t+\beta_{x}^{(i)}\kappa_{t}^{(i)}+\nu_{x,t}^{(i)},\nn\\
& \kappa_t=\varphi_1+\varphi_2t+\rho[\kappa_{t-1}-\varphi_1-\varphi_2(t-1)]+\epsilon_t, \label{model}\\ 
&\kappa_t^{(i)}=\varphi_1^{(i)}+\varphi_2^{(i)}t+\rho^{(i)}[\kappa_{t-1}^{(i)}-\varphi_1^{(i)}-\varphi_2^{(i)}(t-1)]+\epsilon^{(i)}_t, \nn
\end{align}
where $\epsilon_t\stackrel{i.i.d.}{\sim} N(0, \sigma^2_{\kappa})$, $\epsilon_t^{(i)}\stackrel{i.i.d.}{\sim} N(0, \sigma^2_{\kappa^{(i)}})$, $i=1,2,\dots,n$, $x\in\Theta_{\age}$, and $t\in\Theta_{\tim}$. Note that the first line can be viewed as a generalization of (\ref{eq3}) to a multi-population problem, and the last two equations describe the dependence structures of $\kappa_t$ and $\kappa_t^{(i)}$. 
Let $$\bm{U}_{N\times N}= \begin{bmatrix}
1 & 0 & \cdots & \cdots & 0\\ 
-\rho & 1 &  &  & \vdots \\ 
0 & -\rho & \ddots &  & \vdots \\ 
\vdots & \ddots & \ddots & \ddots &  \\ 
0 & \cdots &  & -\rho & 1\\ 
\end{bmatrix} ,
\bm{U}^{(i)}_{N\times N}= \begin{bmatrix}
1 & 0 & \cdots & \cdots & 0\\ 
-\rho^{(i)} & 1 &  &  & \vdots \\ 
0 & -\rho^{(i)} & \ddots &  & \vdots \\ 
\vdots & \ddots & \ddots & \ddots &  \\ 
0 & \cdots &  & -\rho^{(i)} & 1\\ 
\end{bmatrix},
\bm{W}= \begin{bmatrix}
1& t_{1} \\ 
\vdots & \vdots  \\ 
1 & t_{N}\\ 
\end{bmatrix},$$
and also define $\bm{\varphi}=(\varphi_1, \varphi_2)'$, $\bm{\varphi}^{(i)}=(\varphi_1^{(i)}, \varphi_2^{(i)})'$, $\bm{Q=U'U}$, and $\bm{Q}^{(i)}=\bm{U}^{(i)'}\bm{U}^{(i)}$. Then,
$n+1$ dependence structures of time effects can be written as 
\begin{align}
\bkappa \sim N(\bm{W\varphi}, \sigma_{\kappa}^2\bm{Q}^{-1}),\label{dt1}
\end{align} and 
\begin{align}
\bkappa^{(i)} \sim N(\bm{W\varphi}^{(i)}, \sigma_{\kappa^{(i)}}^2(\bm{Q}^{(i)})^{-1}),\label{dt2}
\end{align}
where $\bkappa=(\kappa_1,\kappa_2,\dots,\kappa_N)'$ and $\bkappa^{(i)}=(\kappa^{(i)}_1,\kappa^{(i)}_2,\dots,\kappa^{(i)}_N)'$.

We further assign a dirac spike function to regularize $\varphi_1^{(i)}$ and $\varphi_2^{(i)}$ as follows 
\begin{align}
\varphi_l^{(i)}\sim w_l^{(i)}N(0,c_l^{(i)}\sigma_{\kappa^{(i)}}^2)+(1-w_l^{(i)})\delta_l^{(i)} \label{spike}
\end{align}
 where $w_l^{(i)}$ is a binary variable equal to 1 when a more complicated dependence structure of $\bkappa^{(i)}$ is needed for fitting the data set, and vice versa, $\delta_l^{(i)}$ is a point mass at zero, and $l=1,2$. Note that when all $w_1^{(i)}$s' and $w_2^{(i)}$s' are zero so that $\bm{\varphi}^{(1)}=\bm{\varphi}^{(2)}=\dots=\bm{\varphi}^{(n)}=(0,0)'$, the dependence structures are the same as \cite{antonio2015bayesian}. Although \cite{antonio2015bayesian} justified this special setting in some way, we prefer to consider a more general structure and let data speak out the truth of $\varphi_1^{(i)}$s' and $\varphi_2^{(i)}$s'. With (\ref{spike}), our model can explore the model space of $2^{2n}$ possible dependence structures of $\bkappa^{(i)}$ in a single analysis simultaneously obtaining parameter estimation. When $n$ is big, it can ease computation in model selection compared to using the criteria-based approaches, such as the marginal likelihood criterion and the Akaike information criterion. 
 
 We also want to point out that with the same constraints as used in \cite{antonio2015bayesian},
\begin{align*}
&\sum_{x\in \Theta_{\age}}\beta_x=1,\\
&\sum_{t\in \Theta_{\tim}}\kappa_t=0,\\
&\left\|\beta_{x}^{(i)}\right\|_2 =1,\\
&\sum_{t\in\Theta_{\tim}}\kappa_{t}^{(i)}=0,
\end{align*}
the interpretation of each parameter in (\ref{model}) 
is similar to the one in \cite{antonio2015bayesian}. However, due to the existence of $\nu_{x,t}^{(i)}$, $\alpha_{x}^{(i)}$ can only approximate the mean of log rates at age $x$ across time in the $i^{th}$ population. 
See \cite{antonio2015bayesian} in details for the advantages of such a constraint setting.

\subsection{Prior Specifications}
\subsubsection{Prior Distributions for Age Parameters}
To assure the tractable full conditional distribution of $\alpha_x^{(i)}$, we conduct the same variable transformation $e_x^{(i)}=\exp(\alpha_x^{(i)})$ as  \cite{czado2005bayesian} and \cite{antonio2015bayesian} and propose 
\begin{equation}
 e_x^{(i)} \sim \mbox{Gamma}(a_x^{(i)},b_x^{(i)}),
 \end{equation}
 with the corresponding density $$\pi(e_x^{(i)})=\frac{(b_x^{(i)})^{a_x^{(i)}}}{\Gamma(a_x^{(i)})}(e_x^{(i)})^{a_x^{(i)}-1}\exp(-e_x^{(i)}b_x^{(i)}),$$ where $a_x^{(i)}$ and $b_x^{(i)}$ are pre-specified constants.
 As for $\bbeta=(\beta_1,\beta_2,\dots,\beta_M)'$ and $\bbeta^{(i)}=(\beta^{(i)}_1,\beta_2^{(i)},\dots,\beta^{(i)}_M)'$, we consider the following non-informative priors  
\begin{align*}
&\bbeta \sim N\left(\frac{1}{M}\bm{J}_M, \sigma_{\beta}^2\bm{I}_M\right),\\
&\bbeta^{(i)} \sim N\left(\frac{1}{M}\bm{J}_M, \sigma_{\beta^{(i)}}^2\bm{I}_M\right),
\end{align*}
 where $\bm{J}_M$ is a $M \times 1$ vector with all elements equal to 1, and $\bm{I}_M$ is an identity matrix of size $M$. The hyperparameters $\sigma_{\beta}^2$ and $\sigma_{\beta^{(i)}}^2$ are assumed to follow the inverse Gamma distributions, that is,
\begin{align*}
&\sigma_{\beta}^{2} \sim \mbox{InvGamma}(a_\beta,b_\beta),\\
&\sigma_{\beta^{(i)}}^2 \sim \mbox{InvGamma}(a_{\beta}^{(i)},b_{\beta}^{(i)}),
\end{align*}
where $a_\beta$, $b_\beta$, $a_{\beta}^{(i)}$, and $b_{\beta}^{(i)}$ are pre-specified constants such that $$\pi(\sigma_{\beta}^2)=\frac{b_\beta^{a_\beta}}{\Gamma(a_\beta)}(\sigma_{\beta}^2)^{-a_\beta-1}\exp(-b_\beta/\sigma_{\beta}^2)$$ and $$\pi(\sigma_{\beta^{(i)}}^2)=\frac{(b^{(i)}_\beta)^{a^{(i)}_\beta}}{\Gamma(a^{(i)}_\beta)}(\sigma_{\beta^{(i)}}^2)^{-a^{(i)}_\beta-1}\exp(-b^{(i)}_\beta/\sigma_{\beta^{(i)}}^2).$$ Note that the proposed priors are non-informative in the sense that they are all centered at $1/M$, the constraint (=1) equally shared by $M$ age groups.

\subsubsection{Prior Distributions for Time Parameters}
We consider the following priors for the parameters in (\ref{dt1}) and (\ref{dt2})
\begin{align*}
&\bm{\varphi} \sim  N_2(\bm{\varphi}_0, \bSigma_0),\\
&\rho \sim N(0,\sigma_{\rho}^2)\bm{\mbox{I}}\left\{\rho\in (-1,1)\right\},\\
&\sigma_{\kappa}^{2} \sim \mbox{InvGamma}(a_\kappa,b_\kappa),
\end{align*}
where $\bm{\varphi}_0$, $\bSigma_0$, $\sigma_{\rho}^2$, $a_\kappa$, and $b_\kappa$ are pre-specified hyperparameters, and $\bm{\mbox{I}}\left\{\rho\in (-1,1)\right\}$ is an indicator function equal to 1 when $\rho$ is between -1 and 1. For the dependence structure of $\bkappa^{(i)}$, we propose a conjugate prior for $p^{(i)}\equiv P(w_l^{(i)}=1)$
$$p^{(i)}\sim \mbox{Beta}(a,b),$$
where $a$ and $b$ are pre-specified constants such that $$\pi(p^{(i)})=\frac{\Gamma(a+b)}{\Gamma(a)\Gamma(b)}(p^{(i)})^{a-1}(1-p^{(i)})^{b-1},$$ 
and consider  
\begin{align*}
&\rho^{(i)} \sim N(0,\sigma_{\rho^{(i)}}^2)\bm{\mbox{I}}\left\{\rho^{(i)}\in (-1,1)\right\},\\
&\sigma_{\kappa^{(i)}}^2 \sim \mbox{InvGamma}(a_{\kappa}^{(i)},b_{\kappa}^{(i)}),
\end{align*}
where $\sigma_{\rho^{(i)}}^2$, $a^{(i)}_\kappa$, and $b^{(i)}_\kappa$ are pre-specified hyperparameters.

\subsubsection{Prior Distributions for Overdispersion Parameters} 
Last, following the practical purpose as mentioned in \cite{gelman2006prior}, we assign an Inverse Gamma distribution for $\sigma_{i}^2$
\begin{equation}
\sigma_{i}^2 \sim \mbox{InvGamma}(a_{\mu}^{(i)},b_{\mu}^{(i)}),
\end{equation}
where $a_{\mu}^{(i)}$ and $b_{\mu}^{(i)}$ are pre-specified.

\subsection{Posterior Computation}
\subsubsection{Posterior Distributions for Age Parameters}
Let $\btheta=(e^{(1)},e^{(2)},\dots,e^{(n)}, \bbeta', (\bbeta^{(1)})',(\bbeta^{(2)})',\dots,(\bbeta^{(n)})',\sigma_\beta^2,\sigma_{\beta^{(1)}}^2, \sigma_{\beta^{(2)}}^2,\dots,\sigma_{\beta^{(n)}}^2,$ $\bkappa',\\ (\bkappa^{(1)})', (\bkappa^{(2)})',\dots, (\bkappa^{(n)})',\bvarphi', (\bvarphi^{(1)})', (\bvarphi^{(2)})', \dots, (\bvarphi^{(n)})',\rho,$ $\rho^{(1)},\rho^{(2)},\dots,\rho^{(n)}, \sigma_\kappa^2, \sigma^2_{\kappa^{(1)}},  \\\sigma^2_{\kappa^{(2)}}, \dots, \sigma^2_{\kappa^{(n)}},w^{(1)}_1,w^{(2)}_1,\dots,w^{(n)}_1, w^{(1)}_2,w^{(2)}_2,\dots,w^{(n)}_2, p^{(1)},p^{(2)},\dots,p^{(n)}, \sigma^2_1,\sigma^2_2,\dots,\sigma^2_n,\\\nu_{x,t}^{(1)},\nu_{x,t}^{(2)},\dots,\nu_{x,t}^{(n)})'$. The full conditional distributions of age parameters are given by 
\begin{align}
&\pi(e_x^{(i)}|.)  \propto \exp(-c_x^{(i)}e_x^{(i)})(e_x^{(i)})^{D_{x,.}^{(i)}} \left\vert \frac{d}{de_x^{(i)}} g^{-1}(\alpha_x^{(i)})\right\vert \pi(e_x^{(i)})\nn\\
&\hspace{1.4cm}\propto \exp\left[-(b_x^{(i)}+c_x^{(i)})e_x^{(i)}\right] (e_x^{(i)})^{a_x^{(i)}+D_{x,.}^{(i)}-1}, \label{fc1}\\
&\pi(\beta_x|.)  \propto  \prod_{i=1}^{n}\prod_{t\in \Theta_{\tim}} \exp\left[-E_{x,t}^{(i)}\exp(\alpha_{x}^{(i)}+\beta_x\kappa_t+\beta_{x}^{(i)}\kappa_{t}^{(i)}+\nu_{x,t}^{(i)})\right]\times \exp(\beta_x\kappa_tD_{x,t}^{(i)}) \nn\\
&\hspace{4.4cm} \times \exp\left[-\dfrac{(\beta_x-\frac{1}{M})^2}{2\sigma_\beta^2}\right],\label{fc2}\\
&\pi(\beta_{x}^{(i)}|.)  \propto  \prod_{t\in \Theta_{\tim}} \exp\left[-E_{x,t}^{(i)}\exp(\alpha_{x}^{(i)}+\beta_x\kappa_t+\beta_{x}^{(i)}\kappa_{t}^{(i)}+\nu_{x,t}^{(i)})\right]\times \exp(\beta_{x}^{(i)}\kappa_{t}^{(i)}D_{x,t}^{(i)}) \nn \\
&\hspace{4cm}\times \exp\left[-\dfrac{(\beta_{x}^{(i)}-\frac{1}{M})^2}{2\sigma_{\beta^{(i)}}^2}\right],\label{fc3}
\end{align}
and
\begin{align}
&\pi(\sigma_\beta^2 |.) \propto  (\sigma_{\beta}^2)^{-\tilde{a}_\beta-1}\exp(-\tilde{b}_\beta/\sigma_{\beta}^2),\label{fc4}\\
&\pi(\sigma_{\beta^{(i)}}^2 |.) \propto (\sigma_{\beta^{(i)}}^2)^{-\tilde{a}^{(i)}_\beta-1}\exp(-\tilde{b}^{(i)}_\beta/\sigma_{\beta^{(i)}}^2),\label{fc5} 
\end{align}
where the notation $``|."$ represents ``conditional on all other parameters and the data $G$", $c_{x}^{(i)} = \sum_{t\in \Theta_{\tim}}E_{x,t}^{(i)}\exp(\beta_x\kappa_t+\beta_{x}^{(i)}\kappa_{t}^{(i)}+\nu_{x,t}^{(i)})$,
$D_{x,.}^{(i)} = \sum_{t\in \Theta_{\tim}}D_{x,t}^{(i)}-1$, $\tilde{a}_{\beta}=a_\beta+\frac{M}{2}$, $\tilde{b}_{\beta}=b_\beta+\frac{1}{2}\left(\bbeta-\frac{1}{M}\bm{J}_M\right)'\left(\bbeta-\frac{1}{M}\bm{J}_M\right)$, $\tilde{a}_{\beta}^{(i)}=a_{\beta}^{(i)}+\frac{M}{2}$, and $\tilde{b}_{\beta}^{(i)}=b_{\beta}^{(i)}+\frac{1}{2}\left({\bbeta}^{(i)}-\frac{1}{M}\bm{J}_M\right)'\left({\bbeta}^{(i)}-\frac{1}{M}\bm{J}_M\right)$. From (\ref{fc1}), (\ref{fc4}), and (\ref{fc5}), we have 
\begin{align}
&e_x^{(i)}|. \sim \mbox{Gamma}(a_{x}^{(i)}+D_{x,.}^{(i)}, b_{x}^{(i)}+c_{x}^{(i)} ),\nn\\
&\sigma_\beta^2 |. \sim \mbox{InvGamma}(\tilde{a}_{\beta}, \tilde{b}_{\beta}),\nn\\
&\sigma_{\beta^{(i)}}^2 |. \sim  \mbox{InvGamma}(\tilde{a}_{\beta}^{(i)}, \tilde{b}_{\beta}^{(i)}),\nn
\end{align}
and thus they can be easily sampled in each iteration. 

Due to unidentifiable kernels in (\ref{fc2}) and (\ref{fc3}), the Metropolis-Hastings (MH) sampling is applied to update $\beta_x$ and $\beta_x^{(i)}$. Let $\beta_x^{[j]}$ denote the $j^{th}$ iteration of $\beta_x$, and let $\btheta\backslash \{\}$ denote all parameters in $\btheta$ except the ones in $\{\}$. Assuming that $\beta_y^{[j-1]}$ for $y\geq x$ and $\btheta^{[j]}\backslash\{\beta^{[j]}_y \mbox{ for } y\geq x\}$ are ready, we consider $\beta_x^{*}\sim N(\beta_x^{[j-1]}, \sigma_x^{2})$ as the proposal distribution, where $\sigma_x^{2}$ is chosen to ensure the acceptance probability between 20\% and 40\%. With this symmetric proposal, the acceptance probability  
$$\varPhi(\beta_x^{[j-1]},\beta_x^{*})= \mbox{min}\left\{1,\dfrac{\pi(\beta_x^*|\{\beta_y^{[j-1]} \mbox{for } y> x\}, \btheta^{[j]}\backslash\{\beta^{[j]}_y \mbox{ for } y\geq x\},G) }{\pi(\beta_x^{[j-1]}|\{\beta_y^{[j-1]} \mbox{for } y> x\}, \btheta^{[j]}\backslash\{\beta^{[j]}_y \mbox{ for } y\geq x\},G)}\right\}$$
is compared with a random value $u$ from the Uniform(0,1) and
\begin{align*}
\beta_x^{[j]} =
  \begin{cases}
    \beta_x^*       & \quad \text{if } u \leq \varPhi(\beta_x^{[j-1]},\beta_x^{*})\\
    \beta_x^{[j-1]}  & \quad \text{o.w.}
  \end{cases}.
\end{align*}
To satisfy the constraint $\sum_{x\in\Theta_\age}\beta_x=1$, we let $\tilde{B}=\sum_{y\leq x}\beta_{y}^{[j]}+\sum_{y>x}B_{y}^{[j-1]}$, and update $$(\beta_{x_1}^{[j]},\dots, \beta_{x}^{[j]},\beta_{x+1}^{[j-1]},\dots,  \beta_{x_M}^{[j-1]})\leftarrow\dfrac{(\beta_{x_1}^{[j]},\dots, \beta_{x}^{[j]},\beta_{x+1}^{[j-1]},\dots,  \beta_{x_M}^{[j-1]})}{\tilde {B}}$$ and
$\bkappa^{[j]}\leftarrow\bkappa^{[j]}\tilde{B}$. We repeat all steps above till $x=x_M$ to obtain the $j^{th}$ iteration of $\bbeta$.
For updating $\beta_{x}^{(i)}$ (and also $\bkappa_t^{(i)}$), the similar steps are implemented except that $\tilde{B}$ is calculated by the $L_2$ norm.

\subsubsection{Posterior Distributions for Time Parameters}
In this section, we separately discuss the sampling algorithms for common and population-specific time parameters because the dependence structure of the latter is further regularized by the dirac spike. Let $\bkappa_{-t}=\bkappa\backslash \{\kappa_t\}= (\kappa_1,\dots,\kappa_{t-1},\kappa_{t+1},\dots,\kappa_{t_N})'$ and $\eta_t=\varphi_1+\varphi_2t$, the full conditional distributions of $\bkappa,\bvarphi,\rho$, and $\sigma^2_\kappa$ are proportional to 
\begin{align}
&\pi(\kappa_t|.)  \propto \prod_{i=1}^{n}\prod_{x\in \Theta_{\age}} \exp\left[-E_{x,t}^{(i)}\exp(\alpha_{x}^{(i)}+\beta_x\kappa_t+\beta_{x}^{(i)}\kappa_{t}^{(i)}+\nu_{x,t}^{(i)})\right] \nn
\\
&\hspace{4.3cm}\times \exp(\beta_x\kappa_tD_{x,t}^{(i)})\times f(\kappa_t|\bkappa_{-t}),\label{fc6}\\	
&\pi(\bvarphi|.) \propto  \exp\left[-\frac{1}{2\sigma_{\kappa}^2}(\bvarphi'(\bSigma^{*})^{-1}\bvarphi-2(\bkappa'\bQ\bW+\sigma^2_\kappa\bvarphi_0'\bSigma_0^{-1})\bvarphi)\right],  \label{fc7}\\
&\pi(\rho|.)  \propto \exp\left[-\frac{1}{2\sigma_{\kappa}^2}\left(a_{\rho}\rho^2+\frac{\sigma^2_\kappa}{\sigma^2_\rho}\rho^2-2b_{\rho}\rho\right)\right]\bm{\mbox{I}}\left\{\rho\in (-1,1)\right\}, \label{fc8} \\
&\pi(\sigma_\kappa^2|.)  \propto (\sigma_\kappa^2)^{-(a_\kappa+N/2)-1} \exp\left[-\frac{1}{\sigma_\kappa^2}\left(b_\kappa+\frac{1}{2}({\bkappa-\bW\bvarphi})'\bQ({\bkappa-\bW\bvarphi})\right)\right], \label{fc9} 
\end{align}
where $f(\kappa_t|\bkappa_{-t})$ are the conditional distribution of $\kappa_t$ based on AR(1) with a drift in (\ref{model}), $\bSigma^{*}= (\bW'\bQ\bW+\sigma_\kappa^2\bSigma_0^{-1})^{-1}$, $a_\rho=\sum_{t=t_2}^{t_N}(\kappa_{t-1}-\eta_{t-1})^2$, and $b_\rho=\sum_{t=t_2}^{t_N}(\kappa_t-\eta_t)(\kappa_{t-1}-\eta_{t-1})$.
Note that when $t=t_1$, 
\begin{equation} 
\begin{aligned}
f(\kappa_t|\bkappa_{-t}) &\propto f(\kappa_{t}) f(\kappa_{t+1}|\kappa_t) 
\\
&\propto \exp \left[-\frac{1}{2\sigma_{\kappa}^2}[(\kappa_t-\eta_{t})^2+(\kappa_{t+1}-\eta_{t+1}-\rho (\kappa_t-\eta_{t}))^2]\right];\label{cd1}
\end{aligned}
\end{equation}
when $t_{1}<t<t_N$, 
\begin{align}
f(\kappa_t|\bkappa_{-t}) &\propto f(\kappa_{t+1}|\kappa_t) f(\kappa_{t}|\kappa_{t-1}) \nonumber
\\
&\propto \exp \left[-\frac{1}{2\sigma_{\kappa}^2}[(\kappa_{t}-\eta_{t}-\rho (\kappa_{t-1}-\eta_{t-1}))^2+(\kappa_{t+1}-\eta_{t+1}-\rho (\kappa_t-\eta_{t}))^2]\right];\label{cd2}
\end{align}
and when $t=t_N$, 
\begin{align}
f(\kappa_t|\bkappa_{-t}) \propto f(\kappa_{t}|\kappa_{t-1}) 
\propto \exp \left[-\frac{1}{2\sigma_{\kappa}^2}(\kappa_{t}-\eta_{t}-\rho (\kappa_{t-1}-\eta_{t-1}))^2\right].\label{cd3}
\end{align}
From (\ref{fc7}), (\ref{fc8}), and (\ref{fc9}), $\bvarphi,\rho$ and $\sigma_\kappa^2$ are updated by
\begin{align}
&\bvarphi|.\sim N(\bSigma^*(\bW'\bQ\bkappa+\sigma^2_\kappa\bSigma_0^{-1}\bvarphi_0),\sigma_\kappa^2\bSigma^*),\nn\\
&\rho|.  \sim N\left(\dfrac{b_{\rho}}{a_{\rho}+\frac{\sigma_\kappa^2}{\sigma_\rho^2}},\dfrac{\sigma_{\kappa}^2}{a_{\rho}+\frac{\sigma_\kappa^2}{\sigma_\rho^2}}\right)\bm{\mbox{I}}\left\{\rho\in (-1,1)\right\},\nn\\
&\sigma_\kappa^2|.  \sim \mbox{InvGamma}\left(a_\kappa+\frac{N}{2}, b_\kappa+\frac{1}{2}({\bkappa-\bW\bvarphi})'\bQ({\bkappa-\bW\bvarphi})\right).\nn
\end{align}
To update $\kappa_t$, we let $\kappa_t^{[j]}$ denote the $j^{th}$ iteration of $\kappa_t$ and assume that $\kappa_z^{[j-1]}$ for $z\geq t$ and $\btheta^{[j]}\backslash\{\kappa^{[j]}_z \mbox{ for } z\geq t\}$ are available. Considering $\kappa_t^*\sim N(\kappa_t^{[j-1]}, \sigma_t^{2})$ as the proposal for the MH sampling (similarly, $\sigma_t^{2}$ is selected to have the acceptance probability around $20\%\sim 40\%$), we then have 
\begin{align*}
\kappa_t^{[j]} =
  \begin{cases}
    \kappa_t^*       & \quad \text{if } u \leq \varPhi(\kappa_t^{[j-1]},\kappa_t^{*})\\
    \kappa_t^{[j-1]}  & \quad \text{o.w.}
  \end{cases},
\end{align*}
where 
$$\varPhi(\kappa_t^{[j-1]},\kappa_t^{*})= \mbox{min}\left\{1,\dfrac{\pi(\kappa_t^*|\{\kappa_z^{[j-1]} \mbox{for } z> t\}, \btheta^{[j]}\backslash\{\kappa^{[j]}_z \mbox{ for } z\geq t\}, G) }{\pi(\kappa_t^{[j-1]}|\{\kappa_z^{[j-1]} \mbox{for } z> t\}, \btheta^{[j]}\backslash\{\kappa^{[j]}_z \mbox{ for } z\geq t\},G)}\right\}$$
and $u\sim \mbox{Uniform}(0,1)$. With the constraint $\sum_{t\in\Theta_\tim}\kappa_t=0$, 
$$(\kappa_{t_1}^{[j]},\ldots, \kappa_{t}^{[j]},\kappa_{t+1}^{[j-1]},\ldots,  \kappa_{t_N}^{[j-1]})\leftarrow (\kappa_{t_1}^{[j]},\ldots, \kappa_{t}^{[j]},\kappa_{t+1}^{[j-1]},\ldots,  \kappa_{t_N}^{[j-1]})-\tilde{K}$$ and $(\alpha_x^{(i)})^{[j]}\leftarrow (\alpha_x^{(i)})^{[j]}+\beta_x^{[j]}\tilde{K}$, where $\tilde{K}=(\sum_{z\leq t}\kappa_{z}^{[j]}+\sum_{z>t}\kappa_{z}^{[j-1]})/N$. Repeat all procedures till $t=t_N$ to obtain the $j^{th}$ iteration of $\bkappa$.

To obtain an MCMC sample of $\bkappa^{(i)}$, $\bvarphi^{(i)}$, $\rho^{(i)}$, and $\sigma^2_{\kappa^{(i)}}$, the status of $w_l^{(i)}$ is first determined via $w_l^{(i)}\sim \mbox{Bernoulli}(\xi_l)$. Suppose that $\bm{W}_l$ denotes the $l^{th}$ column of $\bm{W}$ and $\bm{z}^{(i)}_l= \bm{\kappa^{(i)}}-\bm{W}\bvarphi^{(i)}-\bm{W}_{l}\varphi_{l}^{(i)}$, then we have
	\begin{align}
	\xi_l&=\pi(w_l^{(i)}=1| w_{-l}^{(i)}, G)= \frac{\pi(w_l^{(i)}=1| w_{-l}^{(i)}, G)}{\pi(w_l^{(i)}=0| w_{-l}^{(i)}, G)+\pi(w_l^{(i)}=1| w_{-l}^{(i)}, G)}\nn \\
	&= \frac{m(G|w_l^{(i)}=1, w_{-l}^{(i)})\pi(w_l^{(i)}=1)}{m(G|w_l^{(i)}=0, w_{-l}^{(i)})\pi(w_l^{(i)}=0)+m(G|w_l^{(i)}=1, w_{-l}^{(i)})\pi(w_l^{(i)}=1)}\nn \\
	&= \dfrac{p^{(i)}}{p^{(i)}+(1-p^{(i)})R_l^{*}}, \nn
	\end{align}
	where $m(G|w_l^{(i)}=1, w_{-l}^{(i)})\pi(w_l^{(i)}=1)$ and $m(G|w_l^{(i)}=0, w_{-l}^{(i)})\pi(w_l^{(i)}=1)$ are conditional marginal likelihoods measuring the overall model fitting to the data when specifying $w_l^{(i)}$ and $w_{-l}^{(i)}$, and
	\begin{equation}
	R^*_l= \frac{m(G|w_l^{(i)}=0, w_{-l}^{(i)})}{m(G|w_l^{(i)}=1, w_{-l}^{(i)})}
	=(c_l^{(i)} \bm{W}_l'\bm{Q}^{(i)}\bm{W}_l+1)^{1/2}\exp \left[-\dfrac{(\bm{W}_l'\bm{Q}^{(i)}\bm{z}_l^{(i)})^2}{2\sigma_{\kappa^{(i)}}^2(\bm{W}_{l}'\bm{Q}^{(i)}\bm{W}_{l}+1/c_l^{(i)})}\right].
	\end{equation}
Then, based on the identifiable kernels of the full conditional distributions of $\bvarphi^{(i)}$, $\rho^{(i)}$, $\sigma^2_{\kappa^{(i)}}$, and $p^{(i)}$, they are updated by
	\begin{align*}
	&\bvarphi^{(i)}|. \sim N(\bm{a}^*, \bm{A}^*\sigma_{\kappa^{(i)}}^2) \quad \text{if} \quad w_1^{(i)}=w_2^{(i)} =1,\\
	&\rho^{(i)}|. \sim N\left(\dfrac{b_\rho^{(i)}}{a_\rho^{(i)}+\dfrac{\sigma_{\kappa^{(i)}}^2}{\sigma_{\rho^{(i)}}^2}}, \dfrac{\sigma_{\kappa^{(i)}}^2}{a_\rho^{(i)}+\dfrac{\sigma_{\kappa^{(i)}}^2}{\sigma_{\rho^{(i)}}^2}}\right)\bm{\mbox{I}}\left\{\rho^{(i)}\in (-1,1)\right\},\nn\\
	&\sigma_{\kappa^{(i)}}^2|. \sim \mbox{InvGamma}\left(a_k^{(i)}+\frac{N}{2}, b_\kappa^{(i)}+\frac{1}{2}({\bkappa^{(i)}-\bW\bvarphi^{(i)}})'\bQ^{(i)}({\bkappa^{(i)}-\bW\bvarphi^{(i)}})\right),\\		
	&p^{(i)}|. \sim \mbox{Beta}(a+w_1^{(i)}+w_2^{(i)}, b+2-w_1^{(i)}-w_2^{(i)}),
	\end{align*}
	respectively, where $\bm{A}^{*}= (\bW'\bQ^{(i)}\bW+1/c_j^{(i)}\bm{I}_2)^{-1}$, $\bm{a}^{*}= \bm{A}^{*}\bW'\bQ^{(i)}\bkappa^{(i)}$, $\eta_{t}^{(i)}= \varphi_1^{(i)}+\varphi_2^{(i)}t$, $a_\rho^{(i)}= \sum_{t=t_2}^{t_N}(\kappa_{t-1}^{(i)}-\eta_{t-1}^{(i)})^2$, $b_\rho^{(i)}= \sum_{t=t_2}^{t_N} (\kappa_{t}^{(i)}-\eta_{t}^{(i)})(\kappa_{t-1}^{(i)}-\eta_{t-1}^{(i)})$. Note that when $w_1^{(i)}=w_2^{(i)} =0$, $\bvarphi^{(i)}$ is simply updated as $(0,0)'$, and that when $w_{-l}^{(i)}=0, w_{l}^{(i)} =1$, $\varphi_{-l}^{(i)}$ is 0 while $\varphi_{l}^{(i)}$ is updated from the marginal normal distribution with mean and variance equal to the $l^{th}$ and $(l,l)$ elements in $\bm{a}^*$ and $\bm{A}^*\sigma_{\kappa^{(i)}}^2$, respectively. Finally, due to the similar structure of $\pi(\kappa_t^{(i)}|.)$ as $\pi(\kappa_t|.)$
	$$\pi(\kappa_t^{(i)}|.)  \propto \prod_{x\in \Theta_{\age}} \exp\left[-E_{x,t}^{(i)}\exp(\alpha_{x}^{(i)}+\beta_x\kappa_t+\beta_{x}^{(i)}\kappa_{t}^{(i)}+\nu_{x,t}^{(i)})\right] 
\times \exp(\beta_{x}^{(i)}\kappa_{t}^{(i)}D_{x,t}^{(i)}) \times f(\kappa_{t}^{(i)}|\kappa_{-t}^{(i)}),$$
the procedure to update $\kappa_{t}^{(i)}$ is similar to the one for $\kappa_{t}$.

\subsubsection{Posterior Distributions for Overdispersion Parameters}
Because
\begin{align}
\pi(\sigma_{i}^2 |.)\propto & (\sigma_i^2)^{-(a_{\mu}^{(i)}+MN/2)-1} \exp\left[-\frac{1}{\sigma_i^2}\left(b_{\mu}^{(i)}+\frac{1}{2}\sum_{x\in \Theta_{\age}}\sum_{t\in \Theta_{\tim}}(\nu_{x,t}^{(i)})^2\right)\right],\\
\pi(\nu_{x,t}^{(i)}|.)  \propto &  \exp\left[-E_{x,t}^{(i)}\exp(\alpha_{x}^{(i)}+\beta_x\kappa_t+\beta_{x}^{(i)}\kappa_{t}^{(i)}+\nu_{x,t}^{(i)})\right]\times \exp(\nu_{x,t}^{(i)}D_{x,t}^{(i)})\nn\\
 &\times \exp\left[-\frac{(\nu_{x,t}^{(i)})^2}{2\sigma_{i}^2}\right],
\end{align}
we have
\begin{align}
&\sigma_{i}^2 |. \sim \mbox{InvGamma}\left(a_{\mu}^{(i)}+\dfrac{MN}{2}, b_{\mu}^{(i)}+\frac{1}{2}\sum_{x\in \Theta_{\age}}\sum_{t\in \Theta_{\tim}}(\nu_{x,t}^{(i)})^2\right),
\end{align}
and update $\nu_{x,t}^{(i)}$ via the MH sampling. Given that $\btheta^{[j]}\backslash (\nu_{x,t}^{(i)})^{[j]}$ and $(\nu_{x,t}^{(i)})^{[j-1]}$ are available, the proposed function is $(\nu_{x,t}^{(i)})^*\sim N((\nu_{x,t}^{(i)})^{[j-1]}, \sigma_q^{2})$, where $\sigma_q^{2}$ is chosen to have the acceptance probability around 20\%$\sim$40\%. We then have
\begin{align*}
(\nu_{x,t}^{(i)})^{[j]} =
  \begin{cases}
    (\nu_{x,t}^{(i)})^*       & \quad \text{if } u \leq \varPhi((\nu_{x,t}^{(i)})^{[j-1]},(\nu_{x,t}^{(i)})^* )\\
    (\nu_{x,t}^{(i)})^{[j-1]}  & \quad \text{o.w.}
  \end{cases},
\end{align*}
where 
$$\varPhi((\nu_{x,t}^{(i)})^{[j-1]},(\nu_{x,t}^{(i)})^* )= \mbox{min}\left\{1,\dfrac{\pi((\nu_{x,t}^{(i)})^* |\btheta^{[j]}\backslash\{(\nu_{x,t}^{(i)})^{[j]}\}, G) }{\pi((\nu_{x,t}^{(i)})^{[j-1]} |\btheta^{[j]}\backslash\{(\nu_{x,t}^{(i)})^{[j]}\},G)}\right\}.$$

\section{Numerical Analysis} 

\subsection{Data Description}	
The data used to illustrate our proposed method is the Japanese mortality data from the Human Mortality Database (HMD), which contains gender-specific deaths and exposures from 1951 to 2016.
We consider each gender as a single population, and calibrate the model on the data from 1951 to 2000 for ages 0-99 while the data from 2001 to 2016 is separated for the validation purpose. Hence, we have $i=F\mbox{ or }M$, $x_1=0$, $x_M=99$, $t_1=1951$, $t_N=2000$, $M=100$, and $N=50$. 


\subsection{Initial Settings of Prior Distributions}
Following the prior specifications in Sections 3.2.1-3.2.3, we consider $a_x^{(i)}=b_x^{(i)}=1$, $a_{\beta}=b_{\beta}=0.01$, and $a_{\beta}^{(i)}=b_{\beta}^{(i)}=0.001$ as age-related hyperparameters, and set $a_{\mu}^{(i)}=b_{\mu}^{(i)}=2.5$ for overdispersion parameters. For those hyperparameters related to the time factors, they are set as $a=b=1$, $a_{\kappa}=b_{\kappa}=0.001$,
$a_{\kappa}^{(i)}=b_{\kappa}^{(i)}=0.001$, $\sigma_{\rho}^{2}=1$,  $\sigma_{\rho^{(i)}}^{2}=0.1$, $\bvarphi_0=(0,0)^{'}$, $\Sigma_0=\left[
\begin{matrix}
10 & 0 \\
0 & 10 \\
\end{matrix}
\right], $
respectively.
It has to be mentioned that the pre-specified values here are similar to the ones in \cite{czado2005bayesian} and \cite{antonio2015bayesian}, and non-informative relative to the size ($50\times 100\times 2$) of our analyzing data set. 

\subsection{Estimation and Model Performance}
To evaluate the performance of BPLNLCrm, we generate an MCMC sample of 20000 iterations with the first 10000 as burn-ins (so that $j=1,2,\dots,10000$), and summarize the posterior medians of ${\balpha^{(i)}}$, ${\bbeta}$, ${\bbeta^{(i)}}$, ${\bkappa}$ and ${\bkappa^{(i)}}$ along with the 95\% highest posterior density (HPD) intervals in Sections 4.3.1 and 4.3.2, respectively. We also assess the overall model fitting by comparing the posterior predictive distributions of death tolls with the observed counts, and compare the 20-year mortality projection with the method by  \cite{antonio2015bayesian} in Section 4.3.3. For MCMC convergence diagnostics, trace plots of selected parameters are provided in the supplementary materials.
\subsubsection{Estimation for Age Parameters}
\begin{figure}[H]
	\centering  
	\vspace{-0.35cm} 
	\subfigtopskip=1pt 
	\subfigbottomskip=1pt 
	\subfigcapskip=1pt 
	\subfigure[]{
		\label{alphafemale}
		\includegraphics[width=0.84\linewidth]{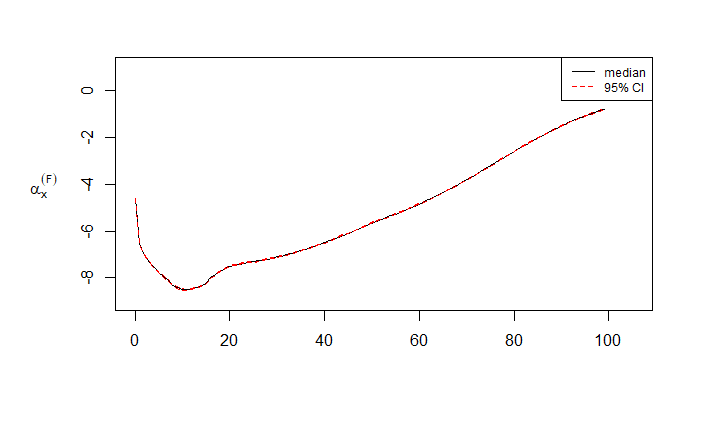}}
	\subfigure[]{
		\label{alphamale}
		\includegraphics[width=0.84\linewidth]{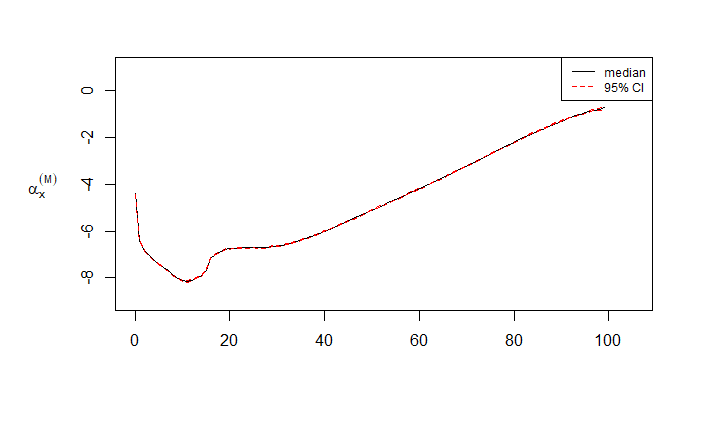}}
	\caption{Plots of the posterior medians of $\alpha_x^{(F)}$ and $\alpha_x^{(M)}$ with their 95\% HDP intervals.}
\end{figure}

\begin{figure}[H]
	\centering 
	\vspace{-0.35cm} 
	\subfigtopskip=1pt 
	\subfigbottomskip=1pt 
	\subfigcapskip=1pt 
	\subfigure[]{
		\label{Beta}
		\includegraphics[width=0.69\linewidth]{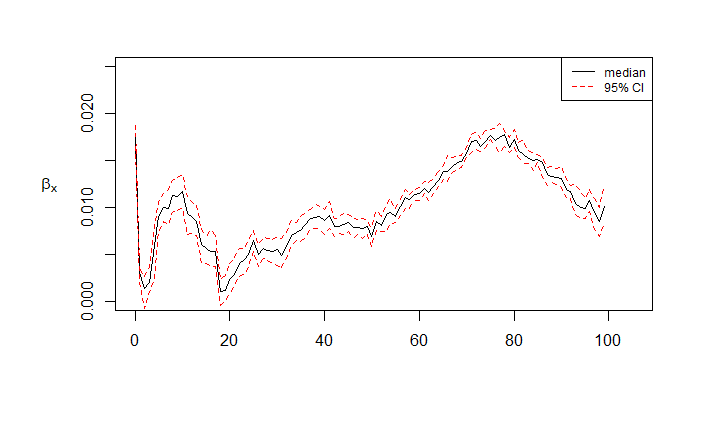}}
	\subfigure[]{
		\label{betafemale}
		\includegraphics[width=0.69\linewidth]{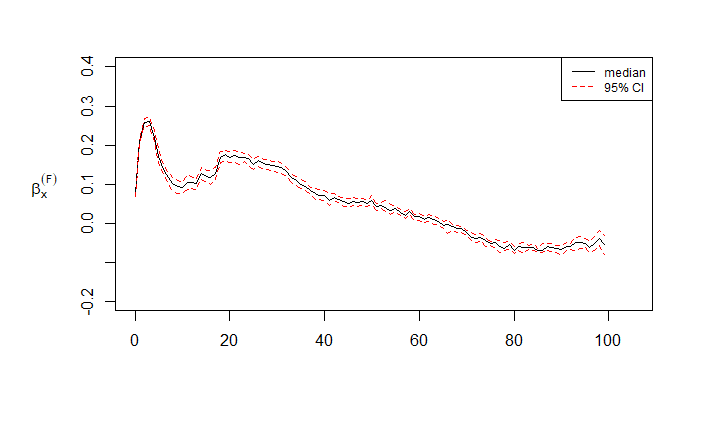}}
	\subfigure[]{
		\label{betamale}
		\includegraphics[width=0.69\linewidth]{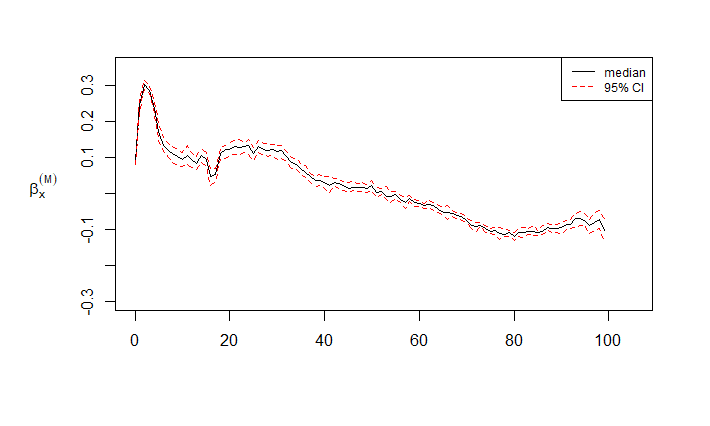}}
	\caption{Plots of the posterior medians of $\beta_x$, $\beta_x^{(F)}$ and $\beta_x^{(M)}$  with their 95\% HDP intervals.}
\end{figure}

Figures \ref{alphafemale} and \ref{alphamale} present the results of $\alpha_x^{(F)}$ and $\alpha_x^{(M)}$ under the BPLNLCrm model, 
where the 95\% HPD intervals are obtained by the method suggested in \cite{Hoff:2009:FCB:1717329}. From Figure 1, we notice that the posterior distributions of  $\alpha_x^{(F)}$ and $\alpha_x^{(M)}$ have small variances, and that there are some features in the estimated curves: First, male tends to have a higher mortality rate than woman. This justifies our multi-population modeling in this case. Secondly, the decline from the infant stage to teenager is likely related to the immune system strengthened with growing age. Then, when ages are around 16-21, the health condition may not be the only decisive factor for the hump. It might be blamed on unnatural deaths caused by immature behaviors in this rebellious stage such as alcohol and drug uses, crimes, and careless drivings etc. For the adult-and-elder stage, the curves consistently go up since deaths happening in this stage are more related to aging. 
Similarly, Figure \ref{Beta} shows the posterior median and 95\% HPD interval of common factor $\beta_x$ while Figures \ref{betafemale} and \ref{betamale} are for the population-specific parameters $\beta_{x}^{(F)}$ and $\beta_{x}^{(M)}$, respectively. It can be seen that the corresponding posterior distributions are concentrated, indicating that the effect sizes of $\beta_x$ and $\beta_{x}^{(i)}$ are less sensitive to all time change at any ages.  
   
\subsubsection{Estimation for Time Parameters}
	Before looking at the results of $\bkappa$, $\bkappa^{(F)}$, and $\bkappa^{(M)}$, we first discuss the selected models for the dependence structures of gender-specific time parameters. Out of 10000 MCMC iterations, we observe that the non-zero proportions of $w_1^{(F)}$, $w_2^{(F)}$, $w_1^{(M)}$, and $w_2^{(M)}$ are 0.96, 0.21, 0.95, and 0.19, respectively. Accordingly, using 50\% as a threshold, the time series models for $\bkappa^{(F)}$ and $\bkappa^{(M)}$ in (\ref{model}) reduce to
\begin{equation}
\kappa_t^{(F)}=\varphi_1^{(F)}+\rho^{(F)}(\kappa_{t-1}^{(F)}-\varphi_1^{(F)})+\epsilon^{(F)}_t,\nn
\end{equation}
and 
\begin{equation}
\kappa_t^{(M)}=\varphi_1^{(M)}+\rho^{(M)}(\kappa_{t-1}^{(M)}-\varphi_1^{(M)})+\epsilon^{(M)}_t,\nn
\end{equation}
respectively.
 	
\begin{figure}[H] 
	\centering  
	\vspace{-0.35cm} 
	\subfigtopskip=1pt 
	\subfigbottomskip=1pt 
	\subfigcapskip=1pt 
	\subfigure[]{
		\label{Kappa}
		\includegraphics[width=0.64\linewidth]{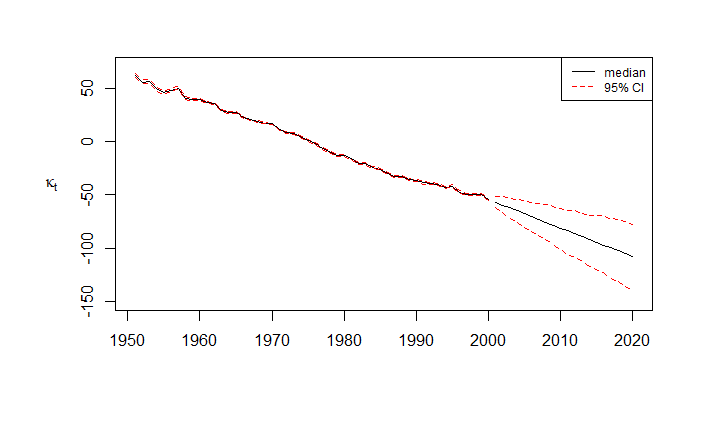}}
	\subfigure[]{
		\label{kappafemale}
		\includegraphics[width=0.64\linewidth]{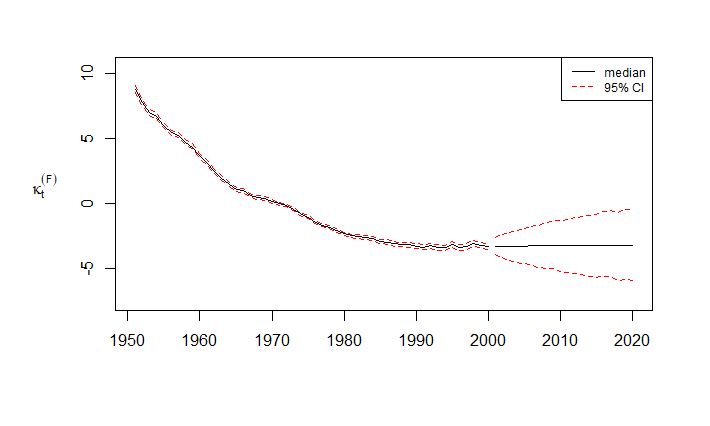}}
	\subfigure[]{
		\label{kappamale}
		\includegraphics[width=0.64\linewidth]{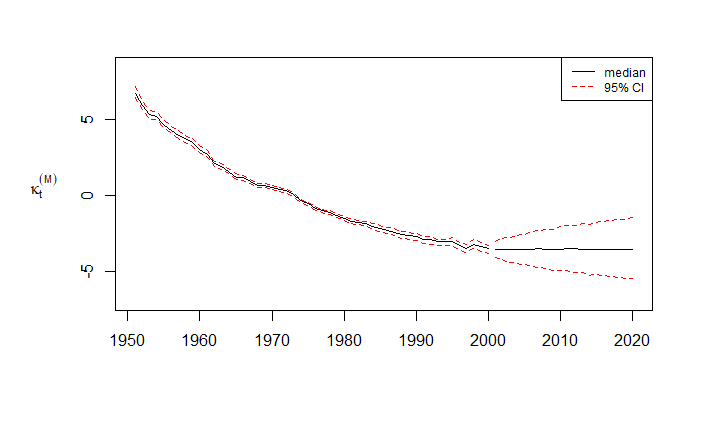}}
	\caption{Plots of the posterior medians of $\kappa_t$, $\kappa_t^{(F)}$ and $\kappa_t^{(M)}$  with their 95\% HDP intervals and the corresponding 20-year ahead projection.}
\end{figure}

In Figures \ref{Kappa}-\ref{kappamale}, we present the posterior medians and 95\% HPD intervals of $\kappa_t$, $\kappa_t^{(F)}$, and $\kappa_{t}^{(M)}$, respectively. In addition to the years 1951-2000, the follow-up 20-year ahead projections of time effects are also provided via the posterior predictive distributions. To obtain a sample from the posterior predictive distribution of $\kappa_t$, we exploit the second equation in (\ref{model}) iteratively. Specifically, we have 
\begin{align*}
\kappa^{[j]}_{N+t'} \sim N\left(\varphi^{[j]}_1+\varphi^{[j]}_2(N+t')+\rho^{[j]}[\kappa^{[j]}_{N+t'-1}-\varphi^{[j]}_1-\varphi^{[j]}_2(N+t'-1)], (\sigma_{\kappa}^2)^{[j]}\right)
\end{align*}
for $j=1,2,\dots, 10000$ and $t'=1,2,\dots,20$. A similar procedure on the third equation in (\ref{model}) is implemented for $\kappa_t^{(i)}$. From Figure 3, we observe decreasing trends in most of time windows, which might be attributed to the advances in medical technology and social welfare. We also observe that $\kappa_t^{(F)}$ and $\kappa_{t}^{(M)}$ have similar estimated curves and will converge to the same size when time passes, meaning that the gender-specific time effects will reach a stable status in the long run as mentioned in  \cite{li2005coherent}.

\subsubsection{Assessments of Model Fitting and Prediction}
The way to empirically peek the posterior distributions of future time effects (that is, $\kappa_{2001},\kappa_{2002},\dots,\kappa_{2020}$) in the previous section can also be used to assess the overall model fitting and its prediction ability. 
Particularly, we compare the observed death tolls $D_{x,t}^{(i)}$ and mortality rates $\mu_{x,t}^{(i)}$ with the ones $(D_{x,t}^{(i)})^{[j]}$ and $(\mu_{x,t}^{(i)})^{[j]}$ that are simulated from the fitted model, where $(D_{x,t}^{(i)})^{[j]}$ follows the Poisson distribution with mean equal to $E_{x,t}^{(i)}(\mu_{x,t}^{(i)})^{[j]}$, and
$(\mu_{x,t}^{(i)})^{[j]}= \exp\left[(\alpha_x^{(i)})^{[j]}+\beta_x^{[j]}\kappa_t^{[j]}+(\beta_{x}^{(i)})^{[j]}(\kappa_{t}^{(i)})^{[j]}+(\nu_{x,t}^{(i)})^{[j]}\right]$ for $j=1,2,\dots,10000$. 

\begin{figure}[H]
	\centering  
	\vspace{-0.35cm} 
	\subfigtopskip=1pt 
	\subfigbottomskip=1pt 
	\subfigcapskip=1pt 
	\subfigure[]{
		\label{female1960}
		\includegraphics[width=0.84\linewidth]{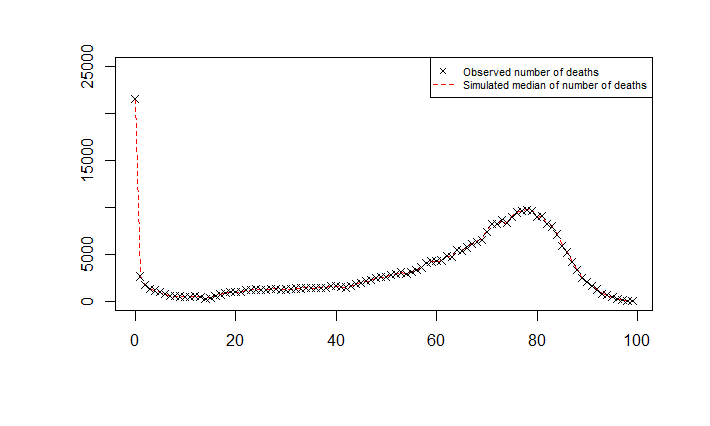}}
	\subfigure[]{
		\label{male1960}
		\includegraphics[width=0.84\linewidth]{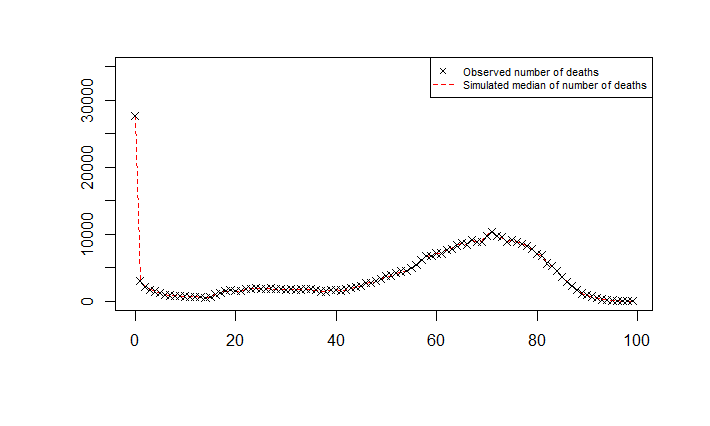}}
	\caption{Plots of the observed and simulated number of deaths
		for (a) female and (b) male in 1960.}
	\label {simu 1960}
\end{figure}

\begin{figure}[H]
	\centering  
	\vspace{-0.35cm} 
	\subfigtopskip=1pt 
	\subfigbottomskip=1pt 
	\subfigcapskip=1pt 
	\subfigure[]{
		\label{female1980}
		\includegraphics[width=0.84\linewidth]{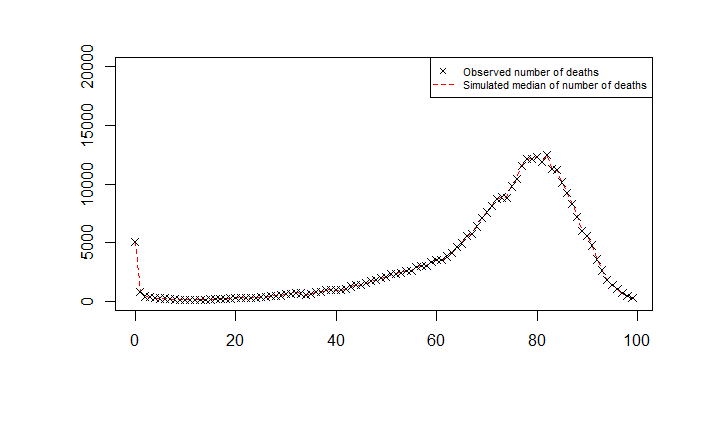}}
	\subfigure[]{
		\label{male1980}
		\includegraphics[width=0.84\linewidth]{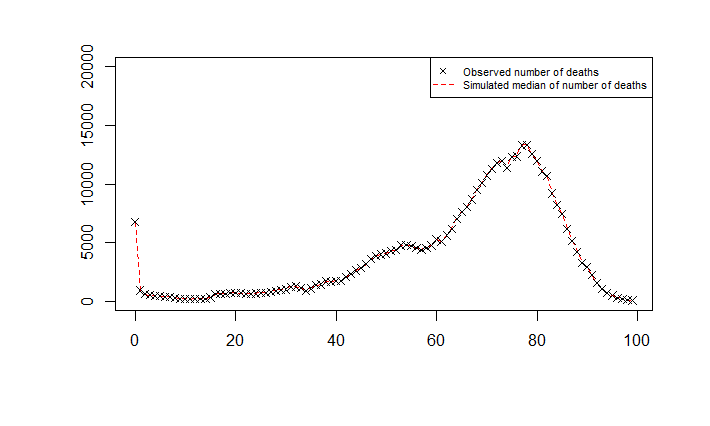}}
	\caption{Plots of the observed and simulated number of deaths
		for (a) female and (b) male in 1980.}
	\label {simu 1980}
\end{figure}

\begin{figure}[H]
	\centering  
	\vspace{-0.35cm} 
	\subfigtopskip=1pt 
	\subfigbottomskip=1pt 
	\subfigcapskip=1pt 
	\subfigure[]{
		\label{female2000}
		\includegraphics[width=0.84\linewidth]{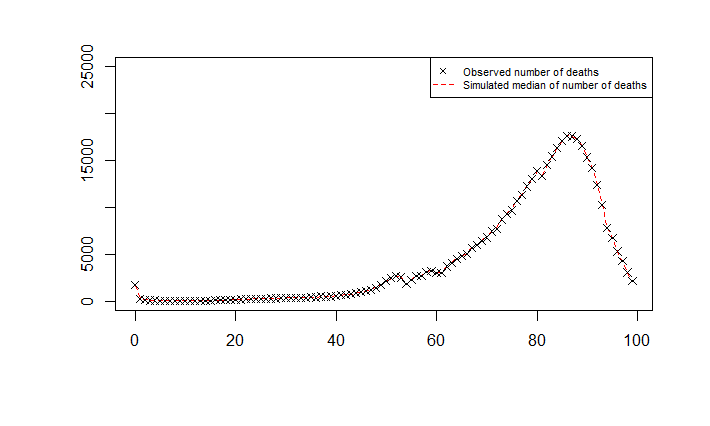}}
	\subfigure[]{
		\label{male2000}
		\includegraphics[width=0.84\linewidth]{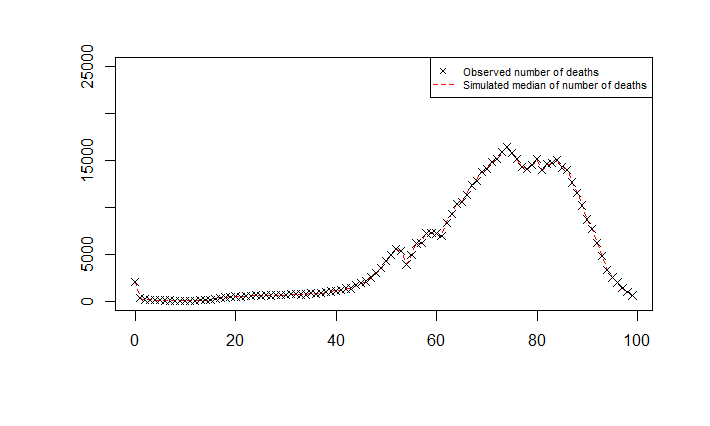}}
	\caption{Plots of the observed and simulated number of deaths
		for (a) female and (b) male in 2000.}
	 \label {simu 2000}
\end{figure}

Figures \ref{simu 1960}-\ref{simu 2000} present the medians of simulated number of deaths for each gender at any ages in three selected years 1960, 1980, and 2000, respectively. It is clear that the simulated values are close to the observed ones, implying that our proposed model fits well to the Japanese mortality data.



\begin{figure}[H]
	\centering  
	\vspace{-0.35cm} 
	\subfigtopskip=1pt 
	\subfigbottomskip=1pt 
	\subfigcapskip=1pt 
	\subfigure[]{
		\label{femaleage15}
		\includegraphics[width=0.84\linewidth]{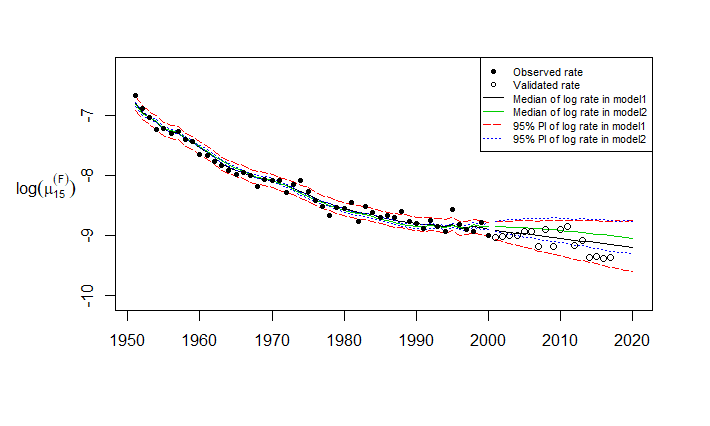}}
	\subfigure[]{
		\label{maleage15}
		\includegraphics[width=0.84\linewidth]{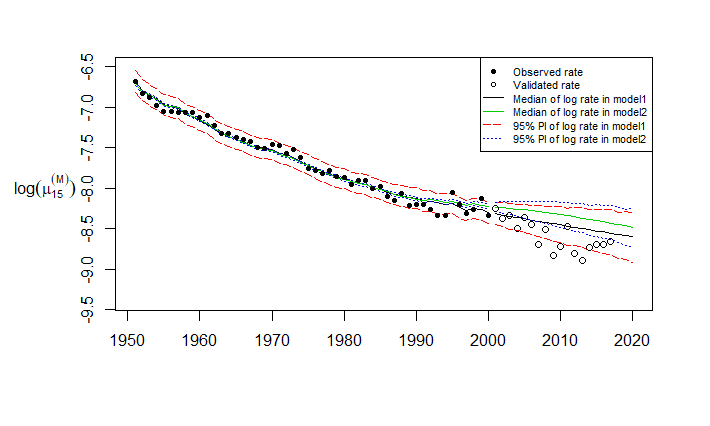}}
	\caption{Plots of the observed and simulated log death rates at age 15 along with 20-year ahead projections and 95\% HDP intervals for (a) female and (b) male.}
	\label{mort 15}
\end{figure}

\begin{figure}[H]
	\centering  
	\vspace{-0.35cm} 
	\subfigtopskip=1pt 
	\subfigbottomskip=1pt 
	\subfigcapskip=1pt 
	\subfigure[]{
		\label{femaleage55}
		\includegraphics[width=0.84\linewidth]{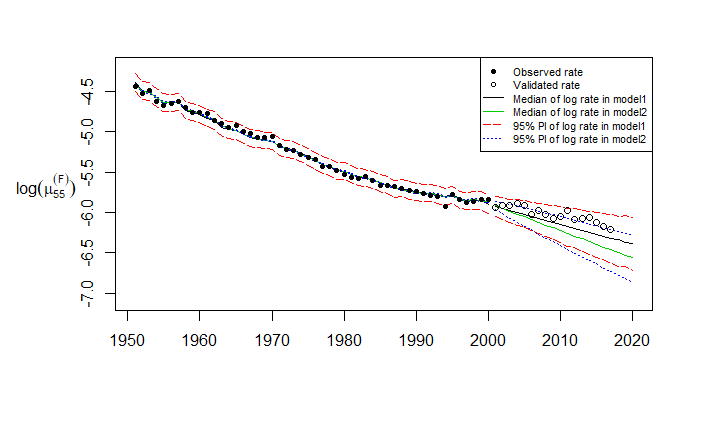}}
	\subfigure[]{
		\label{maleage55}
		\includegraphics[width=0.84\linewidth]{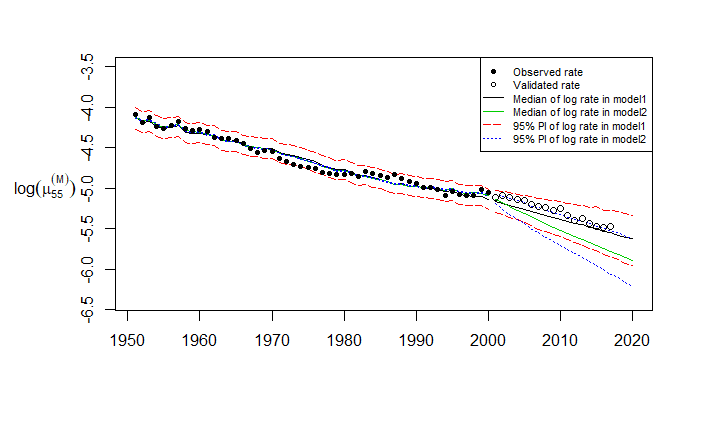}}
	\caption{Plots of the observed and simulated log death rates  at age 55 along with 20-year ahead projections and 95\% HDP intervals for (a) female and (b) male.}
	\label{mort 55}
\end{figure}

\begin{figure}[H]
	\centering  
	\vspace{-0.1cm} 
	\subfigtopskip=0.5pt 
	\subfigbottomskip=0.5pt 
	\subfigcapskip=0.5pt 
	\subfigure[]{
		
		\label{femaleage75}
		\includegraphics[width=0.84\linewidth]{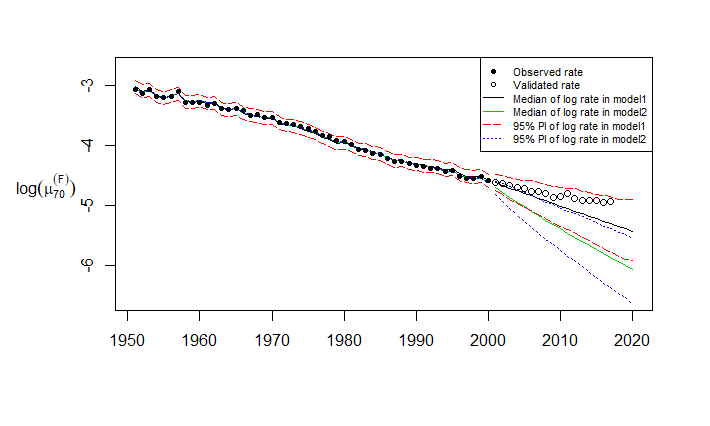}
		
	} 	
	\subfigure[]{
		\label{maleage75}
		\includegraphics[width=0.84\linewidth]{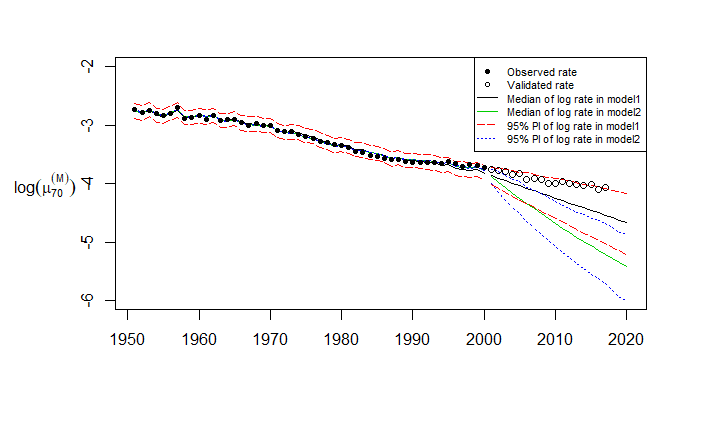}
	}		
	\caption{Plots of the observed and simulated log death rates  at age 70 along with 20-year ahead projections and 95\% HDP intervals for (a) female and (b) male.} 
	\label{mort 70}
\end{figure}

Figures \ref{mort 15}-\ref{mort 70} present the medians and 95\% HPD intervals of simulated log mortality rates at three selected ages 15, 55, and 70, respectively. In addition to the training time window (years 1950-2000), 20-year ahead projections are provided to assess the prediction ability of BPLNLCrm (marked as ``model 1"). We also include the simulated results of the method by \cite{antonio2015bayesian} (marked as ``model 2") as a comparison. 

From Figures \ref{mort 15}-\ref{mort 70}, we observe that the estimated curves (black and green) are close to each other within the training time window, but become bifurcating in the validation. Overall, the BPLNLCrm models provides better 20-years ahead projections because validated log rates are closer to the black curves. We also notice that although the model by \cite{antonio2015bayesian} tends to produce shorter credible intervals (blue dashed curves), those intervals fail to contain many of observed and validated log rates, implying the potential underestimation of variability inherited in model 2. In contrast, the wider credible intervals based on BPLNLCrm, which contain reasonable number of points, may properly present the variability of data by introducing additional overdispersion term, and are preferred.


\newpage
\section{Discussion}
This paper presents a Bayesian approach to estimate and predict mortality for multiple populations under Poisson log-normal assumption. It combines the model by \cite{antonio2015bayesian} with PLNLC, granting the new model to properly reflect the variations of mortality in a multi-population problem. Additionally, by introducing a dirac spike function, the new model can simultaneously conduct model selection and estimation of population-specific time effects. As a result, with this affordable computation even when $n$ is big, it can avoid unnecessary assumptions on dependence structures of $\kappa_t^{(i)}$. It is worth mentioning that as a future work, a more complicated dependence structure can be considered together with the dirac spike function to provide a wider family to explore.  

Another direction to improve BPLNLCrm is to consider the dependence structure of death tolls among populations. In our proposed method, the number of death in each population is treated as conditional independence. However, say using the Japanese mortality data as an example, it is reasonable to believe there are some unmeasurable factors such as culture and dietary habits affecting both female and male mortality. Similarly, when multi-population refers to multi-country, interactions among countries can also be hard to measure and quantify. Hence, the assumption of conditional independence may not be viable and make sense.

\bibliographystyle{plainnat}  
\bibliography{leecar} 	

\end{document}